\documentclass[aps,pre,twocolumn,showpacs,groupedaddress]{revtex4}
\newcommand{\W}{7.5cm}
\usepackage[dvips]{graphicx}
\usepackage{amssymb,latexsym,pifont}
\input epsf

\begin{document}

\title{Influence of the disorder on  tracer dispersion in a flow channel}

\author{V. J. Charette $^{1}$, E. Evangelista
$^{1}$,  R. Chertcoff $^{1}$, H. Auradou $^{2}$, J.~P. Hulin
$^{2}$ and I. Ippolito $^{1}$}

\affiliation{$^{1}$Grupo de Medios Porosos, Facultad de
Ingenier\'ia, Universidad de Buenos Aires, Paseo Col\'on 850, 1063
Buenos Aires, Argentina.}

\affiliation{$^{2}$Laboratoire Fluides, Automatique et Syst\`emes
Thermiques, UMR No. 7608, CNRS, Universit\'es Paris 6 and 11,
B\^atiment 502, Campus Paris Sud, 91405 Orsay Cedex,
France.}

\date{\today}

\begin{abstract}
Tracer dispersion is studied experimentally  in periodic or disordered 
arrays of beads in a capillary tube. Dispersion
is measured  from light absorption variations near the
outlet following a steplike injection of dye at the inlet. Visualizations 
using dye and pure glycerol  are also performed in similar geometries. 
Taylor dispersion  is dominant both in an empty tube and for a periodic
array of beads: the dispersivity $l_d$ increases with 
the P\'eclet number $Pe$ respectively as $Pe$ and $Pe^{0.82}$ and is
larger by a factor of $8$ in the second case. In a disordered packing 
of smaller beads ($1/3$ of the
tube diameter) geometrical dispersion associated to the disorder of the flow
field is dominant with a constant value of $l_d$ reached at high P\'eclet 
numbers. The minimum dispersivity is slightly higher than in homogeneous 
nonconsolidated packings of small grains, likely due heterogeneities resulting from wall effects.
In a disordered packing with the same beads as in the periodic 
configuration, $l_d$ is up to $20$ times lower
 than in the latter and varies as $Pe^\alpha$ with  $\alpha = 0.5$ or $= 0.69$ 
 (depending on the  fluid viscosity). A simple model accounting for this latter 
 result is suggested.

\end{abstract}


\maketitle

\def\nl{\hfill\break}
\parindent= 15pt

\section{Introduction}
\subsection{Objectives of the paper}
Understanding mass and solute transfer in porous and fractured media is relevant to many fields
of science and engineering due to their applications in domains such as waste managament and hydrology~\cite{Bear88,Dullien91,Bear93,Sahimi95}. 
A particularly sensitive method for characterizing mass transfer 
and detecting flow heterogeneities is  the dispersion of a tracer in  a flow
of fluid  through these media.

 In homogeneous $3D$ porous media, the variation in the flow direction $x$ of the concentration $C$ of passive tracer  is often observed  to satisfy at a macroscopic scale (a few pore sizes) the classical convection-diffusion equation~\cite{Bear88}:
\begin{equation}
\frac{\partial C}{\partial t} + U \frac{\partial C}{\partial x}= D_\parallel \frac{\partial^2 C}{\partial x^2} 
\label{eq:condiff}
\end{equation}
Here, $U$ is the flow velocity and $D_\parallel$ is the dispersion coefficient measured in the flow direction and which characterizes the longitudinal spreading of the tracers during their transport.\\
In porous media, the dispersion coefficient is observed to be velocity dependent~\cite{Bear93}. At a low velocity, molecular diffusion predominates and the dispersion coefficient is close to  the molecular diffusion coefficient $D_m$. When the velocity increases, different regimes are observed and non trivial relations such as power law variations of the dispersion coefficient with the velocity are reported~\cite{Seymour97,Maier00,Ippolito94,Bruderer01,Detwiler00,Boschan06}.  
The objective of the present paper is to report dispersion experiments on well controlled models consisting of beads placed inside a long capillary. Different ordered and disordered bead layouts are considered and a rich variety of dispersion behaviors is observed. 
\subsection{Key dispersion mechanisms}
Two physical mechanisms cause the dispersion of passive tracers: molecular diffusion across and along streamlines and advection. Advection refers to tracer transport  by the motion of the host fluid. In order to compare both mechanisms, it is useful to define the P{\'e}clet number: $Pe\ =\ {Ud}/{D_m}$ where $d$ is a characteristic length scale. In this work, $d$ is either the bead or the tube diameter (the latter is used when the flow tube does not contain any beads).

In porous media, the void structure is complex, resulting in tortuous streamlines along which the velocity may greatly fluctuate. In randomly packed beads, for instance, current tubes often get split so that the fluid may flow around a bead; this results in variations of both the magnitude and the direction of 
the local velocity~\cite{Wegner71,Magnico03}. With respect to the average flow, tracers therefore perform a random walk between pores characterized by the duration $t \sim d/U$ of single steps where $d$ is the bead diameter. Using this description, one expects a longitudinal dispersion coefficient $D_{\parallel}/D_m \sim d^2/D_m t \sim Pe$ and a dispersivity, $l_d = D_{\parallel}/U$, constant and corresponding to the correlation length of the velocity field. Yet, regions of slow moving fluid, such as boundary layers, may also play a significant part and lead to logarithmic corrections~\cite{Koch85}.
  
In some specific configurations such as flow in a capillary tube~\cite{Taylor53,Aris56}, in a periodic square array of beads~\cite{Salles93,Maier00} or in a fracture~\cite{Ippolito94,Detwiler00,Boschan06}, the velocity field has long range correlations. In such media, the flow velocity difference between the walls bounding the open space and its center stretches the tracer front and creates concentration gradients: The latter are balanced by molecular diffusion across the gradient. In such a case, the typical time for the decorrelation of the tracer velocity is : $t \sim d^2/D_m$ {\it i.e.} the characteristic time of molecular diffusion across the streamlines and the dispersion coefficient scales like $ D_{\parallel}/D_m \propto Pe^2$. This regime is often called {\it Taylor dispersion} and, for flow in a capillary tube of inner diameter $D$, one finds~\cite{Aris56}:
\begin{equation}
\frac{D_{\parallel}}{D_m} = 1 + \frac{Pe^2}{192},
\label{eq:taylor}
\end{equation}
where $Pe = UD/D_m$. Equivalently, one can write : 
\begin{equation}
\frac{l_d}{d} = \frac{1}{Pe} + \frac{Pe}{192}.
\label{eq:ldtaylor}
\end{equation}
While periodic  $3D$ porous media are very difficult to realize experimentally (even using perfectly
monodisperse beads), the transition from Taylor to geometrical dispersion could
be demonstrated numerically on $2D$ networks of increasing  disorder~\cite{Bruderer01}. 

Yet, some experiments on geological materials, either at the lab or field scales, display non Fickian characteristics. In these situations, the breakthrough curves observed are characterized by tails at long times. This effect may be explain either by the presence of heterogeneities at the scale of the sample or by the trapping of solute in microscopic dead ends or stagnant pores.   

In this work, dispersion experiments in bead packings of varying degrees of disorder are reported. Depending on the structure of the packings, a variety of dispersion regimes is observed ranging from geometrical to Taylor dispersion and including intermediate regimes and non Fickian dispersion.  From these experiments, the tube/bead diameter ratio appears as a key control parameter of the dispersion regime.
\section{Experimental set-up and procedure}
\begin{figure}[h!]
\includegraphics[width=\W]{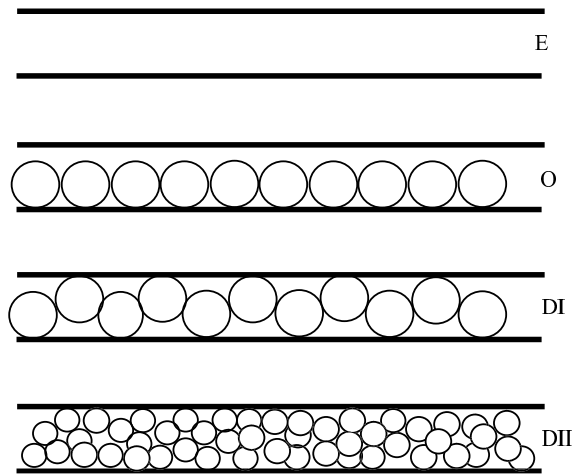}
\caption{Flow geometries used in the experiments. From top to bottom : Empty Channel - Ordered Channel - Disordered Channel I - Disordered Channel II.}
\label{tubitos}
\end{figure}

The experimental model used in this work closely resembles that used by Baudet and coworkers~\cite{Baudet87}.  In this latter work, tracer dispersion was measured  in a long capillary tube filled with monodisperse beads and with a tube/particle diameter ratio of the order of $1.25$. The layout of the beads was either periodic or disordered. The first configuration was achieved by adding one by one beads in an horizontal capillary tube. Finally, one ends up with a line of beads, each of them being in contact with its two neighbors and with the tube wall. The second layout is obtained by  tilting the capillary tubes while filling them. The beads fall then on top of each other and slide often sideways, resulting in a denser packing. The final packing is random and beads are added in the tube until it is entirely filled.\\
In this work, the inner diameter of the capillary tube is $D = 3.1\pm 0.1$~mm, its length is $1500$~mm and well calibrated stainless steel spheres of diameter $d$ are used. 
Beads of diameter $d = 2.54\pm 0.02$~mm were chosen in order to reproduce the configuration used by Baudet {\it et\ al}~\cite{Baudet87}. 
In addition to these two bead packs, two more configurations were used (see Fig.~\ref{tubitos}). The first one uses the same tube but empty. In the last one, the diameter of the beads is $d=1\pm 0.01$~mm, i.e. three time smaller than the tube diameter, leading to a larger tube-to-particle ratio.
In order to clarify the present paper, the following convention is used :
\begin{itemize}
\item  $(E)$ \emph{Empty
channel}: the capillary tube is devoid of beads. 
\item  $(O)$ \emph{Ordered channel}: the tube contains a periodic line of beads touching each other.
\item $(DI)$ \emph{Disordered channel I}: the beads are still touching each other but build up
 a disordered array. The tube/bead diameter ratio is $1.22$. 
\item $(DII)$ \emph{Disordered channel II}: the arrangement is of the same type as $(DI)$ but the diameter of the beads is smaller $d=1\pm 0.01$~mm. The tube/bead diameter ratio is $3.1$.
\end{itemize}
 In configurations {(O)} and {(DI)}, the diameter of the spheres is slightly smaller 
 than the tube diameter ($d=2.54 \pm 0.02$~mm). In configuration $(DII)$, the  diameter of the beads is three times smaller than that of the tube ($d=1 \pm 0.01$~mm)
Originally, one end of the capillary tube is connected to a syringe pump
allowing to establish a stationary flow of the transparent fluid  through the channel: 
A valve allows  to switch the injection to a dyed solution of identical  properties. Fluid flowing out of the
other end of the tube is weighed by computer controlled scales, 
allowing to measure the flow rate throughout the experiment.

The fluids used are solutions of either $10 \%$ or $70\%$ of glycerol in water and their
dynamic viscosities $\mu$ are respectively $1.3 \times 10^{-3}$~Pa.s and  
$23 \times 10^{-3}$~Pa.s (at $20^o C$). 
Water Blue dye at a concentration of $0.05$~g/l is added to one of the solutions: 
It has been selected  because it is chemically stable and does not modify
the rheological properties of the solution. The molecular diffusion coefficient
$D_m$ of the dye was determined through independent Taylor dispersion 
measurements performed  in a vertical teflon capillary tube. 
One obtains in this way
$D_m = 6.5~10^{-4}$~mm$^{2}$/s for the $10\%$ glycerol solution; for the $70 \%$ 
 solution, the value  $D_m = 3.25~10^{-5}$~mm$^{2}$/s is deduced from the value
 for the first solution by the Stokes-Einstein relation~\cite{Russel95}. 
 In the present experiments, 
the P\'eclet number ranges from $50$ to $10^4$ for the $10\%$ glycerol solution but 
reaches $1.4 \times 10^5$ for $70 \%$ glycerol solutions.
In all the studies the Reynolds number is less than $1$ except for the empty tube 
where $Re$ can reach $10$: Under such conditions, the flow can be considered 
as stationary~\cite{Wegner71}.   

In this work, the variation of the tracer concentration at the outlet with time, 
known as the breakthrough curve, is determined from light absorption by the dye. 
The measurement is realized over a square window of size $0.3$ mm$^2$ located
 at $1450$~mm from the injection, close to the outlet and immediately downstream
  of the last bead of the bed.
In order to reduce optical distorsion induced by the tube curvature, 
the measurement  section is enclosed within a transparent plexiglas cell with flat 
parallel walls. The cell is originally filled with glycerol, a fluid with a refractive 
index close to that of the tube.
The section of the tube inside this cell is inserted between a light panel and 
a $4096$ gray levels CCD camera  (Roper Coolsnap CF). 
The set-up is illuminated by a fluorescent tube placed
on the opposite side from the camera and excited at a high frequency 
to reduce fluctuations. 
For each experiment, $2000$ images are recorded by a computer connected 
to the camera at  time intervals ranging 
from $1$ to $30$~sec depending on the flow rate. Dye concentration values
 are determined quantitatively using calibration measurements realized
independently with the experimental tube saturated
with dye solutions of different known concentrations.
Finally, drifts of the light intensity are measured in a region 
of interest outside the tube;  these measurements are then used during
 the analysis of the images to compensate for the effect of these variations on the
transmitted light intensity in the experimental section.

Using this experimental procedure, breakthrough curves were measured for different flow velocities and for different bead layouts. The results of these global measurements are given and discussed in
 section~\ref{sec:secIII}. 
Moreover, in order to improve interpretations of the breakthrough curves, visualizations realized independently  at a  local scale in a similar configuration were carried out. 
In these latter experiments, a transparent tube 
 of $D=8$~mm inner diameter is filled with glass beads of diameters $6$, $3$ or $2$~mm. 
These beads were chosen so that the corresponding tube/bead diameter ratio is respectively $1.33$, $2.67$ and $4$, close to the values used in the dispersion experiment. 
Originally, the model is saturated with glycerol (viscosity $\simeq 1$~Pa.s) which is displaced by the same fluid but dyed. 
Using a fluid with such a high viscosity (and therefore a low diffusion coefficient) allows
to observe and separate clearly the various flow paths at low Reynolds numbers inside the sample. 

Before describing in detail the various dispersive regimes observed for the different bead layouts, the next section describes the methods used to determine the dispersion coefficients from the breakthrough curves.

\subsection{Analysis of experimental curves}
\begin{figure}[h!]
\includegraphics[width=\W]{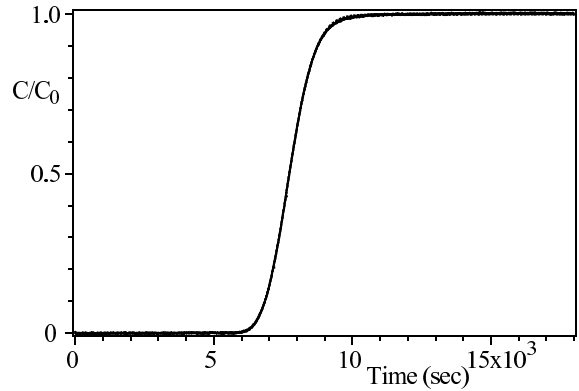}
\caption{Continuous line : typical tracer variation concentration as a function of
time  for a type $DI-$ array and for $Pe=930$ (continuous line).- Dotted line : curve fitted with a variation from   Eq.~(\ref{eq:solucondif}.)}
\label{gaussfit}
\end{figure}

\begin{figure}[h!]
\includegraphics[width=\W]{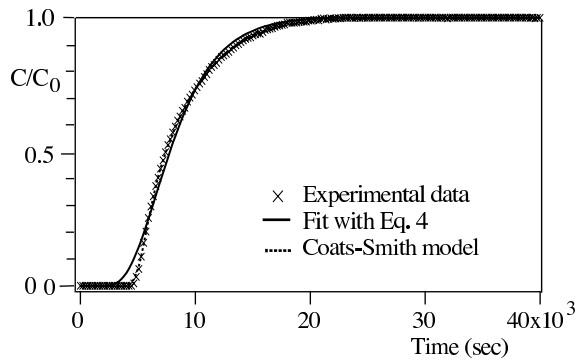}
\caption{Continuous line : experimental dye concentration variation as 
a function of time  for an ordered $O-$type array of
$2.54$~mm diameter  beads and for $Pe=1000$. Dotted lines : 
 Gaussian and Coats-Smith fits.}
\label{asympfit}
\end{figure}

Figure~\ref{gaussfit} displays a typical breakthrough curve obtained after a stepwise  injection of the dyed fluid; asymmetrical curves were also obtained and their analyzis will be discussed later. 
Under such initial conditions and if the concentration satisfies the classical convection-diffusion equation given by Eq.~(\ref{eq:condiff}), the concentration variation at the outlet is given by: 
\begin{equation}
\frac{C(L,t)}{C_0} = \frac{1}{2} (1 - erf\frac{L-U t}{\sqrt{4D_{\parallel}t}}).
\label{eq:solucondif}
\end{equation}
Where, $C_0$ is the dye concentration in the displacing fluid, $L$  the distance between the measurement section and the
 injection point, $D_{\parallel}$ is the longitudinal dispersion coefficient, $U$ is the mean flow velocity. 
The figure~(\ref{gaussfit}) shows the fit of the breakthrough curve by the function given by Eq.~(\ref{eq:solucondif}) where the only adjustable parameter is $D_{\parallel}$. The two curves are almost undistinguishable, indicating that the dispersion process is Fickian and that the classical convection-diffusion  equation~(\ref{eq:condiff}) is satisfied.    
Yet, such perfect fits were not always observed. For instance, in the $O$ configuration and for fairly high velocities $U$, 
one observes at long times, as can be seen on Fig.~\ref{asympfit}, a "tail" effect : it corresponds to an exponential relaxation of the concentration  towards a limiting value with a typical time of the order of $L/U$. 
Such breakthrough curves are better fitted by solutions of the classical Coats-Smith capacitive model~\cite{Coats64} which uses four fitting parameters: $\overline{t}=L/U$, $D_{\parallel}$, $f$ and $t_f$. The two last parameters correspond to the amplitude and the characteristic time of the
exponential variation in the "tail" at long times. By combining these parameters,
one obtains an asymptotic dispersion coefficient $D_{\parallel as}$ :
\begin{equation}
\frac{D_{\parallel as}}{D_m} = \frac{D_{\parallel}}{D_m} + (1 - f)^2U^2t_f,
\label{eq:coatsmith}
\end{equation}
$D_{\parallel as}$ represents the value of the dispersion coefficient that would 
 be measured for channels with the same local structure, but long enough so that a Gaussian
 dispersion regime is reached and Eq.~(\ref{eq:condiff}) becomes valid.
 Note  that the Coats-Smith model is only used here as a mathematical way of 
 obtaining the value of $D_{\parallel as}$; the fact that the experimental curves are 
 well fitted by the model does not mean that its underlying assumptions are valid
 (namely the existence of a zero flow zone with an exponential exchange with
  the mean flow). In the following, the dispersivity $l_d$ is taken equal to $D_{\parallel as}/U$ only for
  experiments  in the $O$ configuration, in the other configurations, $D_{\parallel}$ is obtained by
  fitting the experimental curves to Eq.~(\ref{eq:solucondif}). 

\section{Experimental results}
\label{sec:secIII}
\begin{figure}[h!]
\includegraphics[width=\W]{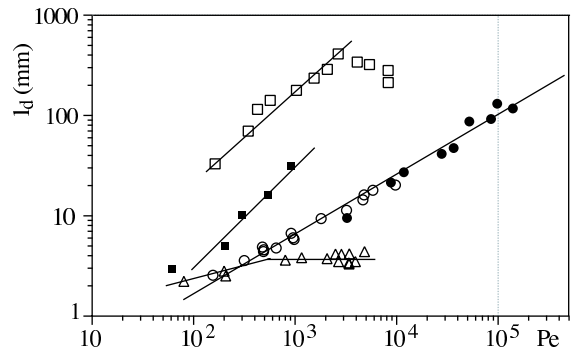}
\caption{Variation of the dispersivity $l_d$ as a function of the
P\'eclet number: ($\vartriangle$) $DII-$array; ($\circ,\bullet$)
$DI-$ array respectively for solutions of $10 \%$ and $70 \%$ glycerol in water;
 ($\blacksquare$) $E-$ array; ($\square$) $O-$
array (in this latter case $l_{d}\ = \ D_{\parallel as}/U$).}
\label{ldvsPe}
\end{figure}

Figure~\ref{ldvsPe} displays variations of the dispersivity
$l_d$ as a function of the P\'eclet number measured for the 
different channel configurations used in the present work.
Note that all the experiments were carried out twice: First, the dyed fluid was injected to displace the clear solution flowing initially in the model at the same flow rate. After the model has been completeley saturated with the dyed fluid, it is displaced in a second experiment by the clear fluid, still at the same flow rate. The dispersion coefficient was found to be identical in the two experiments, implying that no instabilities modified the flow.  
\subsection{Empty tube ($E-$Channel)} 
In the case of an empty tube, all the experimental dispersion curves are well adjusted by the concentration variation given by Eq.~(\ref{eq:solucondif}). 
The corresponding dispersivities $l_d$, as plotted  in Figure \ref{ldvsPe}, increase linearly with $Pe$ reflecting, as expected from Eq.~(\ref{eq:ldtaylor}), a dominant influence of Taylor dispersion. One sees from Eq.~(\ref{eq:ldtaylor}) that the value of $l_d$ is related to the tube diameter; a linear regression of the data gives indeed an effective value of the tube diameter : $a_{eff}\simeq 3 \pm 0.2$~mm.
\subsection{Ordered array of beads ($O-$Channel)} 
Dispersivity $l_{d}$ values  for the ordered array of beads plotted in Figure \ref{ldvsPe}
were obtained with the Coats-Smith model. 
For $Pe \le 2500$, $l_{d}$ varies with $Pe$ 
following a power law : $l_d \sim a_{eff} Pe^\alpha$ (straight 
line in log-log coordinates). From a regression over the 
experimental data, one obtains $a_{eff} \simeq 0.6$~mm and 
$\alpha=0.85\pm 0.2$: This value is compatible
 with a Taylor dispersion mechanism for which $\alpha = 1$.

\begin{figure}[h!]
\includegraphics[width=\W]{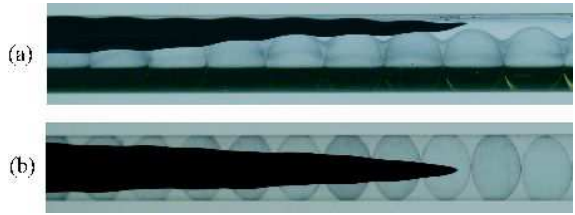}
\caption{Miscible displacement of pure by dyed glycerol in an ordered array of beads inside a capillary tube : (a) side view (b) top view (bead diameter : $d  =  6$~mm; tube diameter : $D  =  8$~mm)}
\label{orden}
\end{figure}

One can indeed expect that, in this channel,  fluid particles follow
streamlines determined by the periodic structure of the bed of beads: Therefore,
as in capillary tubes, tracer can only move from a streamline to
another through molecular diffusion~(See Fig.~\ref{orden}). 
An important feature is the fact that the experimental values of $l_d$ 
which vary from $40$~mm to $400$~mm are about $8$ times larger than for the
empty capillary tube at a same mean flow velocity. This cannot
be explained  by the partial filling up of the flow channel by the
 beads. On the opposite, since these reduce the effective aperture of the flow channels, 
$l_d$ should drop off following Eq.(\ref{eq:ldtaylor}) which is not observed.

Direct visualizations realized in a similar periodic geometry~(Figure~\ref{orden}), 
although with larger beads, help understand these results: Pure glycerol is used
 in these experiments to visualize clearly the boundaries between fluids 
 by removing the influence of molecular diffusion.
In this figure, one observes that the dyed fluid displays a tongue like structure and wraps around the beads. This shape shows that the streamlines are mostly oriented in the flow direction with tiny undulations induced by the beads. Another important point is that the dyed fluid does not flow in the narrow space between the beads: Mass transfert between the two regions can thus only occur through molecular diffusion. This explains the long tail observed in the breakthrough curves. 
Finally, the tongue-like structure reveals the strong velocity contrast between the fluid flow on the side of the beads and above them. 
The velocity gradient stretches the tracer fronts resulting in a concentration gradient which, in turn, gets smoothed by a transverse diffusive flux.\\
All ingredients of a Taylor like dispersion regime are thus present.
Yet, because of the beads, the diffusive flux is more tortuous than in an empty capillary tube: Tracer has to flow around the beads to reach the outlet of the tube. The time needed to homogenize the tracer concentration in the tube section is thus longer, resulting in a higher dispersivity.  

At the highest flow velocities, the variation of $l_{d}$ levels off and it starts
to decrease. A possible explanation is the fact that, in this range of $Pe$ values, the
Reynolds number  becomes higher than $1$ ($Re > 5$): recirculation flows may then develop and induce a more efficient  transverse mixing than molecular diffusion. This
  may account for the decreasing value of $l_d$ at high P\'eclet numbers.

\subsection{Disordered array of small beads ($DII-$array)} 
For a disordered array of $1$~mm diameter beads 
inside the tube, the dispersivity  $l_d$   increases slowly as a function of $Pe$ before reaching a constant value of the order of $3$~mm for $Pe \gtrsim 600$. This constant limiting value of $l_d$ implies that geometrical dispersion associated with the random velocity variations from one pore to the next is dominant. Also, in such geometries, the correlation length of the velocities of fluid particles along their path is too short for Taylor dispersion to have a significant influence. Dispersion characteristics of such arrays are then comparable to those of non consolidated packings of monodisperse grains: In this latter case, the minimum value of $l_d$ is lower than the grain size (typ. $0.6-0.7\,d_g$) and is reached for P\'eclet numbers (based on the grain size) of the order of $10$~\cite{Han85}. In Fig.~\ref{ldvsPe}, the value of $l_d$ for a similar P\'eclet number ($Pe = 30$ when based on the tube diameter) is $l_d = 2$~mm or about twice the bead diameter. This result  also suggests that boundary layers have  a weaker influence on dispersion than predicted by some theories~\cite{Koch85} since no increase of $l_d$ as $Log Pe$ is observed. 

\begin{figure}[h!]
\includegraphics[width=\W]{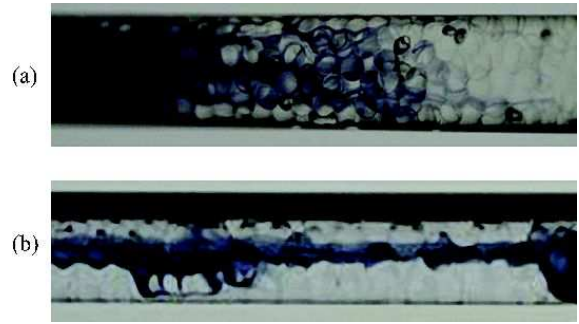}
\caption{Miscible displacements of pure glycerol by dyed glycerol saturating 
disordered glass beads packings inside a $D = 8$~mm diameter capillary tube. Bead diameter : (a) $d = 2$~mm  - (b)  $d = 3$~mm. The tube/particle diameter ratio is respectively $4$ and $2.66$.}
\label{deso23mm}
\end{figure}
Visualizations realized with $d = 2$ and $3$~mm beads inside a $D = 8$~mm diameter tube (Figs~\ref{deso23mm}a-b) help understand these results.
Figure~\ref{deso23mm}a demonstrates clearly the many divisions of the front of injected fluid after moving through several pores and the rather uniform distribution of the invading fluid across the flow section :
for the less viscous fluids used  in our dispersion experiments, transverse molecular diffusion would mix quickly these thin filaments with the surrounding fluid, leading to the Gaussian dispersion observed for the $DII-$ array. 

For the lower value $2.66$ of the ratio $d/D$  the shape of the dye streaks changes very much as  shown on Figure~\ref{deso23mm}b. In this particular case, flow is strongly channelized in the center of the column; this indicates, that for some (low) values of the ratio $d/D$, pathological structures of the flow field may appear. The dispersion measurements reported in the present section were performed with a column for which $d/D$ was intermediate between the two values in Figs.~\ref{deso23mm}a-b. Since no channelization of the flow was observed visually during the measurement,  the geometrical dispersion observed corresponds most likely to the type of flow structure shown in Fig.\ref{deso23mm}a. 
In the next section, tracer dispersion is investigated for still smaller values of $d/D$.
\subsection{Disordered array of large beads ($DI-$array)} 
In this case, the beads have the same diameter as in the periodic $O-$ channel (i.e. $d=2.54$~mm), but the packing is now disordered (See Fig.~\ref{tubitos}). The experiments were performed using the two water-glycerol solutions containing either $10\%$ or $70\%$ of glycerol in weight. 
As for the $DII$ array and in contrast with the ordered one, all breakthrough curves are well adjusted by Eq.~(\ref{eq:solucondif}) allowing for the determination of $D_\parallel$ for the various flow conditions. The Figure~\ref{ldvsPe} shows the dispersivity $l_d=D_\parallel/U$ as a function of the P{\'e}clet number for the two solutions used in the experiments.\\
Clearly, the dispersivity $l_d$ and $Pe$ have a power law relationship. Fitting the variation of $l_d$ with $Pe$ to  $a_{eff} Pe^{\alpha}$ gives respectively for the $10\%$ and $70\%$ glycerol solutions  $\alpha=0.52\pm0.01$, $a_{eff}\simeq 0.2$~mm and $\alpha=0.69\pm0.04$, $a_{eff}\simeq 0.04$~mm. 
One observes that $a_{eff}$ and $\alpha$ are only slightly different for the two solutions  despite a ratio of $20$ between the viscosities. 
One notices that the values of $l_d$ for a given P\'eclet number are much lower for the disordered array than for the periodic one (by a factor of $15$ (resp. $40$  at low (resp. high) $Pe$ values). Also, at low P\'eclet numbers, $l_d$ becomes of the order of the bead size and gets close to the dispersivity observed for the DII array.

\begin{figure}[h!]
\includegraphics[width=\W]{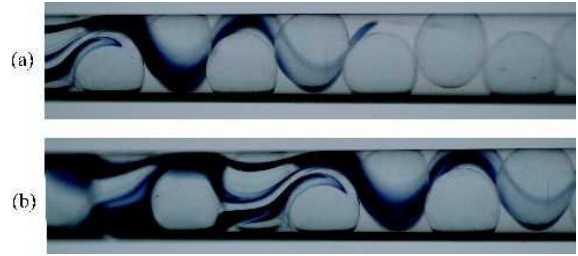}
\caption{Views at different times of the miscible displacement front of pure by dyed glycerol in a disordered channel: the diameter of the beads is $d = 6$~mm
and the diameter of the tube $D = 8$~mm. The tube-to-particle ratio is $1.33$.}
\label{deso1}
\end{figure}

An important feature is the fact that the exponent $\alpha$ characterizing the variation of $l_d$ with $Pe$ is of the order of $0.5$ (for the $10\%$ solution) and of $0.69$ for the other one (at slightly higher P\'eclet numbers). These exponents  are intermediate between the values $0.82 \pm 0.2$ for the periodic array and $1$ for the empty tube and the value $0$ (constant $l_d$) for the disordered packing of smaller beads: This implies that the dispersion mechanism is intermediate between Taylor  and geometrical dispersion corresponding respectively to the first and second cases.
This result was also observed by Baudet {\it et al}~\cite{Baudet87} but remained unexplained. \\
As in the previous sections, visualization on a larger system with the  same $d/D$ ratio help understand the dispersion mechanisms. The visualizations of Figure~\ref{deso1}a-b show that the dyed fluid is mostly split into two streaks  located near the walls. This channelization results from  the increase of the porosity near the walls~\cite{Mueller92};   moreover, the low value of the tube/bead diameter ratio (of the order of $1$)  breaks the angular isotropy of the porous structure and concentrates the flow paths in a few (here $2$) channels. Magnico~\cite{Magnico03} estimated  that, at low Reynolds numbers, a fluid layer of thickness of the order of $d/4$ appears, inside which fluid flow is  purely longitudinal and tangential with no radial component. At first, one expects therefore radial exchange between the dye streaks, clearly visible on Figure \ref{deso1}a, and the remaining pore space to be mostly diffusional. Yet, as can be seen of the Figure \ref{deso1}b structural heterogenety resulting from misplaced beads split the streaks into filaments parallel to the mean flow~\cite{Wegner71}; the number of filaments increases then along the flow path and their size decreases. These filaments persist over distances significantly larger than the bead  diameter so that molecular diffusion may spread the filaments over transverse distances of the order of their size. This gives rise to Taylor-like dispersion so that the global dispersion results from the combined influences of geometrical and Taylor dispersions. Such an influence of the flow channelization on dispersion was recently reported by Bruderer {\it et al}~\cite{Bruderer01} in $2D$ networks.  

\subsection{Qualitative model of different power law variations}
A qualitative argument helps understand how one can reach such power law 
variations of $l_d \propto Pe^\alpha$ ($0 < \alpha < 1$) under the combined influence 
of the disorder of
the flow field and of transverse molecular diffusion. Assume that the front gets divided into 
streaks of width $a_x$ decreasing with the distance $x$ parallel 
to the flow as  $a_{x} \sim d^{1+\beta} x^{-\beta}$ 
(the $d$ term allows to have the right dimensionality for the equation). 
By generalizing the Taylor argument, the transition to diffusive spreading 
should occur when the transverse molecular diffusion time $\tau_{diff}$ across the distance 
$a_x$ is of the order of the mean transit time $L/U$ with :
\begin{equation}
\tau_{diff} \sim \frac{a_x^2}{D_m} \sim \frac{x}{U}
\label{eq:taudif}
\end{equation}
Replacing $a_x$ by its expression provides the distance at which the
transition should take place :
 \begin{equation}
 \, x_{trans} \sim d (\frac {U d}{D_m})^{1/(1+2\beta)} \sim d Pe^{1/(1+2\beta)}
\label{eq:decor}
\end{equation} 
With, as usual, $Pe = Ud/D_m$. As in Taylor dispersion, $x_{trans}$, represents the characteristic decorrelation distance
 of the velocity of tracer particles and we assume therefore that $D_{\parallel} \sim U x_{trans}$
  leading to
\begin{equation}
D_{\parallel eff} \sim Ud Pe^{1/(1+2\beta)}
\label{eq:Dalpha}
\end{equation}
or 
\begin{equation}
\frac{D_{\parallel eff}} {D_m} \sim Pe^\frac{2 + 2\beta}{1+2\beta}
\label{eq:DbetaPe}
\end{equation}
For $\beta\, = \,0$ (no geometrical variation of the filament width with distance), one retrieves Taylor dispersion with $D_{\parallel Taylor } \sim a^2U^2/D_m$ and for $\beta = \infty$ (fast decorrelation), one obtains geometrical dispersion with $D_{\parallel geom} \sim d U$ (the exponent $\alpha$ defined previously should then be related to $\beta$ by $\alpha = 1/(1 + 2\beta)$ 
In the present case, the experimental result $D_{\parallel} \propto Pe^{3/2}$  implies that $\beta = 1/2$. Therefore, the experimental observations on $DI-$ may be accounted for by assuming a combination
of the influences of transverse molecular diffusion and of the geometrical disorder of the packing with, for the latter, a rate of division of the fluid streaks intermediate between those observed in a packing of small beads and in a capillary tube.
\section{Conclusion}
To conclude, despite its simple structure (a long tube filled up with beads), the experimental system studied in  the present work displayed a broad variety dispersion regimes. Both the layout of the beads and the ratio of their diameter to that of the tube were shown to influence very much the dispersion characteristics.\\
First, Taylor dispersion is observed for a periodic bead array inside a tube of slightly larger diameter than the beads, like in the empty tube but with an increased dispersion coefficient. 
With the same beads inside the same tube, but packed in a disordered array, the dispersivity $l_d$ is very strongly reduced and the exponent $\alpha$ characterizing the variation of $l_d$ with $Pe$ decreases from almost $1$ to $0.5$.
This variation reflects the reduced persistence length of fluid streaks as they move along the tube (this length remains  however much larger than the bead diameter). A simple model assuming a power law reduction of the width of the streaks of dye with distance allows to reproduce this variation of $l_d$ with $Pe$.
When the ratio of the tube and particle diameters is increased by using beads of smaller diameter, the dispersion coefficient varies with the P\'eclet number as $D_{\parallel} \propto Pe$ for $Pe\ >\ 600$. This reflects  a geometrical dispersion regime with a short persistence of the dye streaks controlled by the local pore geometry and a weak influence of diffusion in the boundary layers.\\
 More observations of the flow and concentration fields at the local scale (using for instance matched index fluids) are needed to explain \emph{quantitatively} these results. The increase of the dispersivity for the periodic array compared to the empty tube and  the decorrelation of the motions of the fluid particles in the disordered arrays for small tube/particle diameter ratios are two particularly important issues.\\
\section*{Acknowledgments}
We thank J. Koplik and G. Drazer for very helpful comments
and discussions. This work was supported by
UBA-$IN029$, CNRS-Conicet International Cooperation Program (PICS
$N^{\circ}\, 2178$) and by the ECOS Sud program
$N^{\circ}\,A03E02$.

\end{document}